# Temperature Dependence of Magnetophonon Resistance Oscillations in GaAs/AlAs Heterostructures at High Filling Factors

A.A.Bykov, A.V.Goran

*Institute of Semiconductor Physics, 630090 Novosibirsk, Russia*

The temperature dependence of phonon-induced resistance oscillations has been investigated in two-dimensional electron system with moderate mobility at large filling factors at temperature range $T = 7.4 – 25.4$ K. The amplitude of phonon-induced oscillations has been found to be governed by quantum relaxation time which is determined by electron-electron interaction effects. This is in agreement with results recently obtained in ultra-high mobility two-dimensional electron system with low electron density [A. T. Hatke *et al.*, Phys. Rev. Lett. **102**, 086808 (2009)]. The shift of the main maximum of the magnetophonon resistance oscillations to higher magnetic fields with increasing temperature is observed.

The magnetoresistance oscillations caused by electrons interacting with acoustic phonons were discovered [1] in high-mobility two-dimensional electron systems (2DES) at large filling factors (i.e. when condition $E_F/\hbar w_c \gg 1$ is satisfied, where $E_F$ is Fermi energy, $\hbar w_c$ is the distance between Landau levels). These oscillations are periodic in inverse magnetic field and originate from resonance electron scattering on phonons with wave vector equal to double Fermi wave vector and energy equal to $\hbar w_c$. The period of phonon-induced resistance oscillations observed in [1] at large filling factors is determined by the ratio $w_s/w_c = (2k_F)u_s/w_c = j$ where $k_F$ is Fermi wave vector, $u_s$ is sound velocity, $w_c$ is cyclotron frequency and $j$ is positive integer.

The temperature dependence of $w_s/w_c$ oscillations in ultra-high electron mobility 2DES ($m \sim 1.2 \times 10^7$ cm$^2$/Vs) with low electron density ($n_e \sim 3.8 \times 10^{11}$ cm$^{-2}$) was recently studied and the amplitude $Dr_{PIRO}$ of oscillations can be written as [2]:

$$Dr_{PIRO}(T) \propto t_{ph}^{-1}(T)\exp[-2p/w_c t_q^{ee}(T)], \qquad (1)$$

where $t_{ph}$ is the relaxation time due to electron-acoustic phonon scattering mechanism, $t_q^{ee}$ is quantum relaxation time due to electron-electron scattering mechanism. The component $t_{ph}^{-1}(T)$ in Eq. (1) is responsible for the growth of the amplitude of $w_s/w_c$ oscillations at low $T$, and $\exp(-2p/w_c t_q^{ee})$ is responsible for damping of oscillations at higher $T$. The amplitude of $w_s/w_c$ oscillations has its maximum at some optimal temperature $T_0$ which increases with rising of magnetic field $B$. Assuming $1/t_{ph}(T) \propto T^\alpha$ [3-5] and $1/t_q^{ee}(T) = lT^2/E_F$ [6,7] the optimal temperature $T_0$ can be written as:

$$T_0 = k_B^{-1}(aE_F\hbar\omega_c/4pl)^{1/2}, \qquad (2)$$

where $a$ and $l$ are dimensionless constants. The authors of [2] found that $T_0^2$ linearly depends on magnetic field for oscillations numbered $j = 1, 2, 3$: $T_0^2 \propto B$ and suggested that the amplitude of phonon-induced oscillations is governed by quantum relaxation time modified by electron-electron interaction.

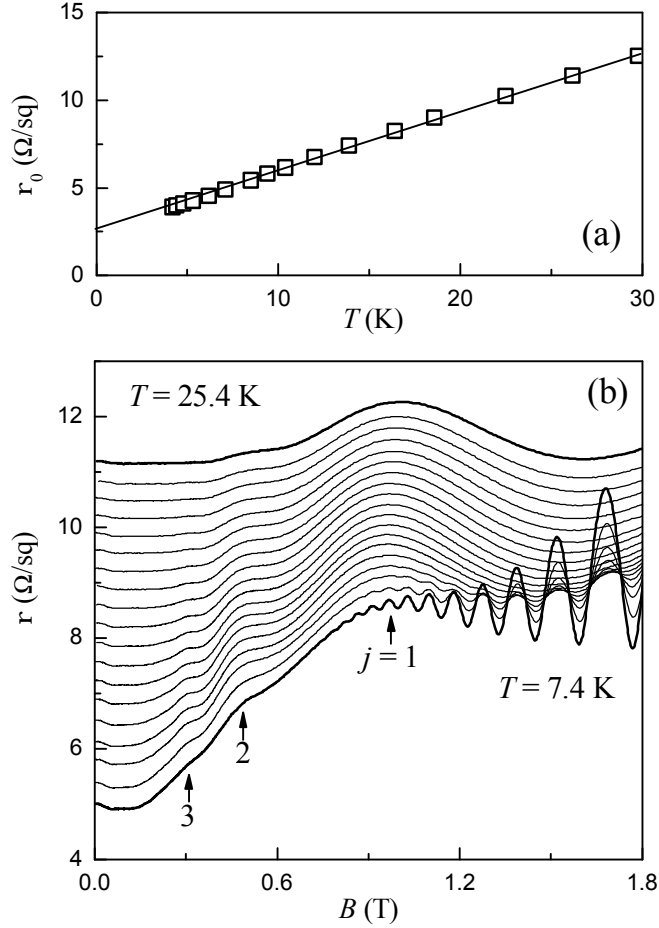

Fig 1. (a) Temperature dependence $r(T)$ of the resistivity of 2D electron gas in GaAs quantum well with AlAs/GaAs superlattice barriers. (b) Resistivity $r(B)$ at different temperatures from 7.4K (lowest curve) to 25.4 K (highest curve) in 1 K steps. Arrows mark the maxima of $w_s/w_c$ oscillations.

In this paper we present an experimental study of the temperature dependence of $w_s/w_c$ resistance oscillations in 2DES with lower mobility ($m \sim 2 \times 10^6$ cm$^2$/Vs) but higher electron density ($n_e \sim 8 \times 10^{11}$ cm$^{-2}$) then 2DES studied in [2]. Despite the lower mobility we could observe $w_s/w_c$ oscillations in wide temperature range, which was enough for studying their temperature dependences. We got experimental data that is in good qualitative agreement with the results of [2] and confirmed the role of electron-electron scattering in temperature damping of $w_s/w_c$ oscillations in 2DES. The shift of the main maximum of the magnetophonon resistance oscillations to higher magnetic fields with increasing temperature is observed.

We used symmetrically doped single GaAs quantum wells (width $w$=13 nm) with AlAs/GaAs superlattice barriers [8,9] grown using molecular-beam epitaxy on (100) GaAs substrates. Magnetoresistance measurements were performed on 450x50 $m$m Hall bars in the temperature range $T = 4.2 - 30$ K and magnetic field $B < 2$ T. Electron density was $n_e = 7.7 \times 10^{11}$ cm$^{-2}$ and zero-field mobility was $m = 1/en_e r_0 = 2.1 \times 10^6$ cm$^2$/Vs at $T$=4.2 K. Resistivity $r$ was measured on low-frequency (777 Hz) current not exceeding $10^{-6}$A.

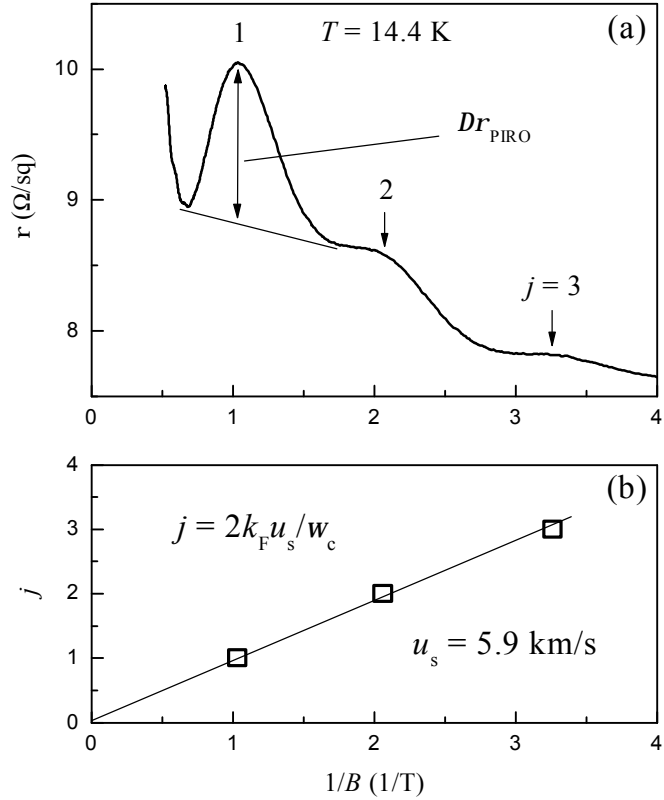

Fig. 2. (a) Resistivity $r(1/B)$ as function of inverse magnetic field at temperature $T = 14.4$ K. Arrows mark the peaks corresponding to $w_s/w_c = 1, 2$ and $3$. (b) Dependence of $j = w_s/w_c = 2k_F u_s/w_c$ vs $1/B$. The streight line corresponds to $u_s = 5.9$ km/s.

The Figure 1a presents the typical $r_0(T)$ dependence in the GaAs/AlAs heterostructures under study. Experimental data is well approximated by the linear function. This means that the constant $a$ in expression $1/t_{ph}(T) \propto T^{\alpha}$ is close to unity [4]. The Figure 1b presents $r(B)$ dependence in the temperature range $T = 7.4 - 25.4$ K. One can see Shubnikov-de Haas oscillations at $T=7.2$ K in magnetic field $B > 1$ T with the inverse-field periodicity corresponding to electron density $n_e$ obtained from the Hall data. Shubnikov-de Haas oscillations disappear at higher temperature, but a new kind of oscillations appears, with peaks numbered 1, 2 and 3. With increasing temperature the amplitude of these oscillations increases until it reaches the maximum at some $T_0$, then it decreases. The peak numbered 3 completely disappears at $T=25.4$ K, while the peak numbered $j=1$ is slightly shifted to higher magnetic fields with increasing $T$.

The Figure 2 illustrates that these oscillations are periodic in inverse magnetic field and can be explained by electron-phonon interaction with sound velocity of $u_s \sim 5.9$ km/s [10,11]. We note that the sound velocity calculated from the period of acoutsic phonon induced magnetoreresistance oscillations ranges from 2.9 to 4.8 km/s in the samples with a lower density [1, 2]. The difference observed in the $u_s$ values for the 2D systems with different values of $n_e$ has recently been theoretically explained by the interaction of electrons with the transverse and longitudinal modes of the bulk acoustic waves in the GaAs quantum wells grown on the (100) surface [12]. However, this theory fails to explain the temperature shift of the main maximum ($j = 1$) of the magnetophonon resistance oscillations experimentally revealed in the GaAs quantum wells with AlAs/GaAs superlattice barriers.

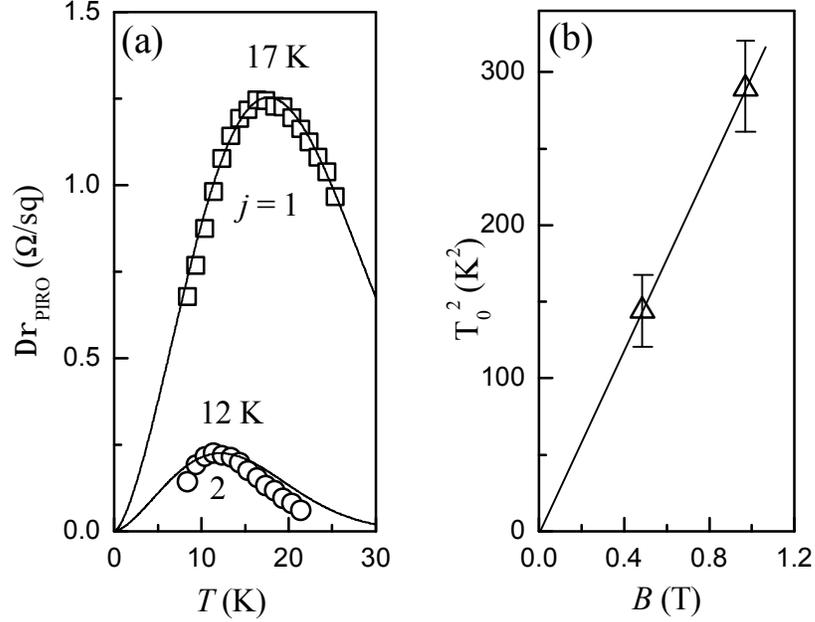

Fig. 3. (a) Dependence of $Dr_{PIRO}$ vs $T$ at $j = 1$ and 2. Curves were calculated using Eq. (1) for $a = 1.5$ and $l = 2.5$. (b) Squared optimal temperature $T_0^2$ vs magnetic field $B$. Linear function corresponds to $a/l = 0.6$.

The Figure 3a presents the temperature dependence of $\omega_s/\omega_c$ oscillations amplitude for $j = 1, 2$. The amplitude exhibits non-monotonic behavior with maxima, which is consistent with Eq. (1). The maximum for $j = 1$ ($T_0 = 17$ K) manifests itself at higher magnetic field than the maximum for $j = 2$ ($T_0 = 12$ K). On the Figure 3b one can see that $T_0^2$ is a linear function of $B$, which is in perfect agreement with Eq. (2). Using Eq. (2) we can calculate the ratio $a/l = 0.6 \pm 0.05$. Assuming that in our electron system $a$ should be about 1 we find that $l$ should be about 1.7. However with values of $a = 1$ and $l = 1.7$ we couldn't obtain a good agreement between $Dr_{PIRO}(T)$ calculated using Eq. (1) and our experimental curves. Good agreement for $j = 1$ could be obtained with values of $a = 1.5$ and $l = 2.5$. At these values of $a$ and $l$, the experimental and calculated curves for $j = 2$ are in worse agreement. The temperature dependence at $j = 3$ has been measured with a low accuracy and is not presented here.

We think that the temperature dependence of the parameter $a$ is responsible for the fact that the temperature dependences of the $\omega_s/\omega_c$ oscillation amplitude at $j = 1$ and 2 cannot be described by the same set of fitting parameters $a$ and $l$. The assumption made in [2] that $1/t_{ph}(T) \propto T^\alpha$, where $a$ is the constant, is not strict because $a$ is generally temperature dependent. In particular, $a$ in [2] varies from 5 to 1.8 in the temperature range of 2–7 K. To obtain an accurate dependence of $1/t_{ph}(T)$ from the $m(T)$ dependence, we should know the temperature independent mobility component $m_{im} = m(T = 0)$ due to the scattering on a random potential [5]. The value of $m_{im}$ cannot be experimentally determined accurately because the zero temperature cannot be experimentally obtained. Figure 4a presents the curve of $1/m(T)$ in the temperature range of 4.2–30 K. This dependence implies that $m_{im}$ ranges from $m_{im}^{min} = 2.1 \times 10^6$ cm²/Vs to $m_{im}^{max} = 3 \times 10^6$ cm²/Vs.

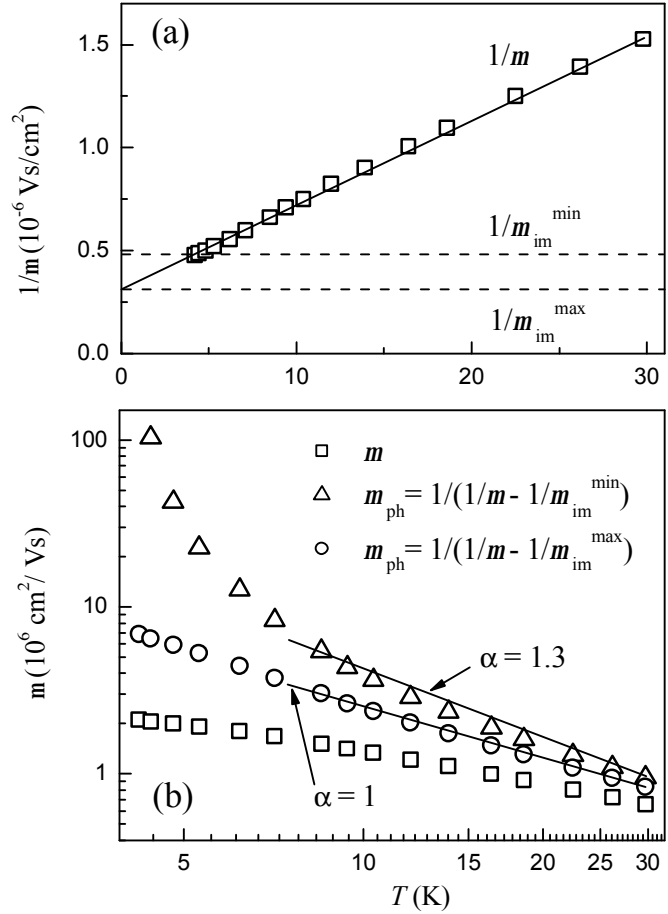

Fig. 4. (a) Temperature dependences of $1/m$ and $1/m_{im}$. The straight line is a linear approximation. The dashed lines are the temperature dependences of $1/m_{im} = 1/m_{im}^{max}$ and $1/m_{im} = 1/m_{im}^{min}$. (b) The temperature dependences of $m$ and $m_{ph}$. The straight lines correspond to the linear approximation at $a = 1$ and 1.3.

In the temperature range under study, the $m(T)$ dependence can be represented as [5]:

$$1/m(T) = 1/m_{im} + 1/m_{ph}(T), \qquad (3)$$

where $m_{ph}$ is the component of $m$ due to the scattering of electrons on acoustic phonons. Figure 4b presents the experimental curve $m(T)$ and the dependences of $m_{ph}(T)$ calculated from this curve by formula (3) for the cases of $m_{im} = m_{im}^{max}$ and $m_{im} = m_{im}^{min}$. If $a = $ const, then the $m_{ph}$ values in the log–log scale should fit the straight line whose slope is determined by $a$. It is seen that this is valid only if $m_{im} = m_{im}^{max}$. However, if $m_{im} = m_{im}^{min}$, then the curve $m_{ph}(T)$ is not described by the relation $m_{ph}(T) \propto 1/T^a$; i.e., in the general case, this approximation is wrong and the dependence of $a(T)$ should be taken into account. In our case, due to a large error in the measurement of $m_{im}$, the $m_{ph}(T)$ curve lies between the $m_{ph}(T)$ curves calculated for the two extreme cases of $m_\text{m} = m_\text{m}^\text{max}$ and $m_\text{m} = m_\text{m}^\text{min}$. Assuming that $1/m_\text{ph}(T) \propto T^a$, where $a$ is the dimensionless constant, in the measurement range of the $Dr_\text{PIRO}(T)$ dependence at $j = 1$, the value of $a$ determined from $m_\text{ph}(T)$ lies between 1 and 1.3. Nevertheless, the main conclusion of this paper that the electron–electron scattering plays the dominant role in the temperature suppression of the amplitude of the acoustic phonon induced magnetic field resistance oscillations remains valid.

In conclusion, we have studied the temperature dependence of $w_\text{s}/w_\text{c}$ oscillations in 2D electron system with moderate mobility. We have found that the amplitude of $w_\text{s}/w_\text{c}$ oscillations exhibits non-monotonic behavior in samples under study, temperature, and the main maximum of these oscillations is shifted to higher magnetic fields with increasing temperature $T$. From experimental curves we calculated the values of the constants: $a = 1.4 \pm 0.1$ and $l = 2.5 \pm 0.2$. Our experimental results are in good qualitative agreement with the results presented in [2] and confirm the important role of electron-electron scattering on transport properties of 2D electron systems [2, 13-17].

We are grateful to A.V. Chaplik, M.V. Entin, and L.I. Magarill for useful discussions. This work was supported by RFBR Project No. 08-02-01051.